\documentclass[%
 reprint,
 amsmath,amssymb,
 aps,
 prl,
]{revtex4-2}

\usepackage{graphicx}
\usepackage{dcolumn}
\usepackage{bm}
\usepackage{xcolor}
\usepackage{ulem}

\usepackage{float}
\restylefloat{table}

\begin{document}

\preprint{APS/123-QED}

\title{Effects of Neural Heterogeneity on Spiking Neural Network Dynamics}

\author{Richard Gast}
 \email{richard.gast@northwestern.edu}
\author{Sara A. Solla}
\author{Ann Kennedy}%
\affiliation{%
 Department of Neuroscience, Feinberg School of Medicine, Northwestern University, Chicago, USA.\\
}%



\date{\today}

\begin{abstract}
The brain is composed of complex networks of interacting neurons that express considerable heterogeneity in their physiology and spiking characteristics.
How does neural heterogeneity affect macroscopic neural dynamics and how does it contribute to neurodynamic functions? 
In this letter, we address these questions by studying the macroscopic dynamics of networks of heterogeneous Izhikevich neurons.
We derive mean-field equations for these networks and examine how heterogeneity in the spiking thresholds of Izhikevich neurons affects the emergent macroscopic dynamics.
Our results suggest that the level of heterogeneity of inhibitory populations controls resonance and hysteresis properties of systems of coupled excitatory and inhibitory neurons.
Neural heterogeneity may thus serve as a means to control the dynamic repertoire of mesoscopic brain circuits.
\end{abstract}

\maketitle


\section{The Study of Neural Heterogeneity}

Neurons exhibit striking heterogeneity in their structure, gene expression, and electrophysiological response properties \cite{izhikevich_simple_2003,harris_neocortical_2015, yao2021taxonomy}.
In cortex, about 50 types of GABAergic interneurons have been identified; many of these have been hypothesized to serve distinct circuit functions \cite{lim_development_2018}. 
However, cell type boundaries in functional, anatomical, or single-cell sequencing datasets are often fuzzy, and neurons of the same cell type can express considerable variance in their structure and response properties. \cite{marsat_neural_2010,angelo_biophysical_2012}.
Thus a more appropriate model of cell type diversity in the brain should reflect the contributions of both true genetically distinct cell types and of heterogeneity across cells within a given type. 
We call the latter "within-type" heterogeneity, in contrast to "between-type" heterogeneity across distinct genetically or functionally defined cell types.

How do within-type and between-type heterogeneity contribute to brain function?
While various theories have been developed regarding the contributions of between-type heterogeneity to brain function and behavior \cite{bastos_canonical_2012,gittis_striatal_2012,hangya_circuit_2014}, the role of within-type heterogeneity has received less attention.
Nonetheless, within-type heterogeneity has been shown to affect encoding and decoding properties of neural populations \cite{shamir_implications_2006,padmanabhan_intrinsic_2010,savard_neural_2011,borges_self-sustained_2020}.
Furthermore, decreases in within-type neural heterogeneity have been suggested as a key factor in the generation of seizure-like oscillations \cite{rich_loss_2022}.

Here we examine the interaction of within-type and between-type heterogeneity in shaping the macroscopic dynamics of populations of interacting spiking neurons.
While previous studies have demonstrated that within-type heterogeneity affects the input-output relationships and resonance properties of specific neural circuits \cite{golomb_dynamics_1993,perez_constructive_2010,mejias_optimal_2012,brito_neuronal_2021,rich_loss_2022}, we aim for a more general characterization of how neural heterogeneity affects the dynamics of circuits containing interconnected neurons of different types.

To do so, we leverage mean-field theory to examine the dynamics of spiking neural networks by making use of a recently developed mean-field theory for networks of coupled spiking neurons with distributed parameters \cite{luke_complete_2013,montbrio_macroscopic_2015}. 
This theory has been used to examine how neural population dynamics change based on cell-intrinsic properties \cite{gast_role_2021,pietras_mesoscopic_2022}, coupling between populations \cite{byrne_mean_2017,dumont_macroscopic_2019,taher_exact_2020}, and mechanisms such as gap junction coupling and synaptic plasticity \cite{pietras_exact_2019,gast_mean-field_2020,taher_bursting_2022}.
Here, we characterize the distinct contributions of within-type and between-type heterogeneity by adapting the mean-field theory to the case of coupled networks containing distinct functional cell types.

\section{Mean-Field Models of Coupled Heterogeneous Spiking Neural Networks}

\subsection{The Spiking Neural Network}

In order to study both within-type and between-type heterogeneity, we model each cell type as a network of coupled Izhikevich (IK) neurons \cite{izhikevich_simple_2003} with parameters tuned to replicate the spiking behavior of a variety of functional cell types. 
This network model takes the form

\begin{align}
    C \dot v_i &= k (v_i - v_r)(v_i - v_{\theta}) - u + I + g s (E-v_i), \label{eq:v_i}\\
    \tau_u \dot u &= b(\frac{1}{N}\sum_{j=1}^N v_j - v_r) - u + \frac{\tau_u \kappa}{N} \sum_{j=1}^N \delta(v_j-v_p), \label{eq:u_i}\\
    \tau_s \dot s &= -s + \frac{J \tau_s}{N} \sum_{j=1}^{N} \delta(v_j-v_p). \label{eq:s}
\end{align}

Here $v_i$ represents the membrane potential of the $i^{th}$ neuron in the network. 
The variable $u$ contains two terms; the left term controls the attractiveness of the resting membrane potential, whereas the right term accounts for spike-frequency adaptation to neural firing, and is controlled by the spiking history of the full population. 
The variable $s$ reflects the low-pass filtering effect of the synapse and dendritic tree integration on the synaptic inputs to a cell. 
The original IK model included a neuron-specific recovery variable $u_i$ \cite{izhikevich_dynamical_2007}; here we consider a simplification based on a global recovery variable $u$ shared across neurons. 
It has recently been show that this global model yields a good approximation to the dynamics of an IK network with neuron-specific recovery variables in the case where $u_i \gg \kappa, \forall i \in {1,2,...,N}$ \cite{chen_exact_2022}.

Note that the spike in the IK model does not occur instantaneously after crossing the threshold $v_{\theta}$; rather, $v_i$ continues to rise until $v_i \geq v_p$; where $v_p$ is the peak membrane potential.
When that condition is satisfied, a spike is counted and $v_i$ is reset to the reset potential $v_0$. 
The total firing rate of the network at time $t$ is given by $r(t) = \frac{1}{N} \sum_{j=1}^N \delta(v_j-v_p)$ where $\delta$ is the Dirac delta function; it appears on the right-hand sides of eqs.\eqref{eq:u_i} and \eqref{eq:s}.
This firing rate drives the synaptic activation $s$, modeled as a convolution of the mean-field firing rate with an exponential activation kernel with decay time constant $\tau_s$; it also drives the recovery variable $u$.
Additional parameters that control the behavior of the network are the cell capacitance $C$, the leakage parameter $k$, the resting and threshold potentials $v_r$ and $v_{\theta}$, the synaptic coupling strength $J$, and the parameters $\tau_u$, $b$, and $\kappa$ that control the time evolution of the recovery variable. 
Finally, additional input currents to the network can be defined via the global input parameter $I$.

To study the effects of within-type heterogeneity on a population of these neurons, we must (a) derive the mean-field equations that govern the network dynamics, and (b) identify relevant sources of heterogeneity.
To address (a), we leverage recent results in mean-field theory: that the population dynamics of some types of spiking neural networks are fully captured by their average firing rate and average membrane potential, and that their mean-field equations can be derived via the Ott-Antonsen ansatz or the equivalent Lorentzian ansatz \cite{ott_low_2008,montbrio_macroscopic_2015,bick_understanding_2020}.
To address (b), we note that neural heterogeneity can be realized by treating model parameters as distributed in value across the population.
For example, mean-field theories of spiking neural networks typically consider the input variable $I$ as a distributed quantity. 
However, the input $I$ only enters into eq.\eqref{eq:v_i} as an additive term. 
In contrast, another parameter that contributes to within-type heterogeneity, the spiking threshold $v_{\theta}$ \cite{wilson_excitatory_1972,rich_loss_2022}, appears in eq.\eqref{eq:v_i}, $v_{\theta}$ as multiplied by the state variable $v_i$, and thus directly couples to the current state of the neuron as characterized by its membrane potential.
Thus, while distributions over $I$ capture heterogeneity in the tonic synaptic drive to a population, distributions over $v_{\theta}$ capture heterogeneity of cell-intrinsic properties within a population.
Here, we will consider $v_{\theta}$ as our heterogeneity-inducing parameter. The values of  $v_{\theta,i}$ within the network are neuron-specific and drawn from a probability distribution $p(v_{\theta})$.

\subsection{Derivation of the Mean-field Equations}

We consider the system given by eqs.(\ref{eq:v_i}-\ref{eq:s}) in the thermodynamic limit $N \rightarrow \infty$.
In this limit, the state of the system can be captured by a density $\rho(v,v_{\theta},t)$: the probability of observing a neuron characterized by a spike threshold $v_{\theta}$ in a state of membrane potential $v$ at time $t$.   
The system dynamics can then be expressed via the continuity equation

\begin{align}
    \frac{\partial}{\partial t} \rho(v,t|v_{\theta}) &= - \frac{\partial}{\partial v} [G^v(v,u,s,v_{\theta}) \rho(v,t|v_{\theta})], \label{eq:cont_cond}\\
     G(v,u,s,v_{\theta}) &= \frac{1}{C} [k (v - v_r)(v - v_{\theta}) - u + I + g s (E-v)]. \label{eq:flux}
\end{align}

Here we used the factorization $\rho(v,v_{\theta},t) = \rho(v,t|v_{\theta}) p(v_{\theta})$ to separate the dynamical variables from the quenched neuron-specific parameter $v_{\theta}$. 

To derive a self-consistent set of equations that captures these dynamics, we apply the Lorentzian ansatz outlined in \cite{montbrio_macroscopic_2015}.
First, we make the assumption that the distribution over $v$ at any time $t$ is fully captured by a Lorentzian probability distribution with center $y$ and half-width-at-half-maximum $x$:

\begin{equation}
    \rho(v,t|v_{\theta}) = \frac{1}{\pi} \frac{x(t,v_{\theta})}{[v-y(t,v_{\theta})]^2 + x(t,v_{\theta})^2}. \label{eq:rho_lorentzian}
\end{equation}

As shown in \cite{montbrio_macroscopic_2015}, $x$ and $y$ are intrinsically related to dynamic quantities: the average firing rate $r = \frac{1}{N}\sum_{j=1}^N r_j$ and the average membrane potential $v = \frac{1}{N}\sum_{j=1}^N v_j$, via

\begin{align}
    r(t) &= \frac{k}{C \pi} \int_{v_{\theta}} x(t,v_{\theta}) p(v_{\theta}) d v_{\theta}, \label{eq:rx}\\
    v(t) &= \int_{v_{\theta}} y(t,v_{\theta}) p(v_{\theta}) d v_{\theta}. \label{eq:vy}
\end{align}

By plugging eq.\eqref{eq:rho_lorentzian} into eq.\eqref{eq:cont_cond} and equating the left- and right-hand-sides in powers of $v$, we find that the system dynamics obey 

\begin{equation}
    C \frac{\partial}{\partial t} z(t,v_{\theta}) = i[-kz(t,v_{\theta})^2 + \alpha z(t,v_{\theta}) + \beta] \label{eq:z}
\end{equation}

in terms of the complex dynamical variable $z(t,v_{\theta}) = x(t,v_{\theta}) + iy(t,v_{\theta})$. Here we have defined $\alpha = k(v_r+v_{\theta}) + g s$ and $\beta = k v_r v_{\theta} + g s E - u + I$.
Finally, to derive the equations for $r(t)$ and $v(t)$ from eq.\eqref{eq:z}, we need to solve the integral

\begin{equation}
    \dot z = \frac{1}{C} \int_{v_{\theta}} \frac{\partial}{\partial t} z(t,v_{\theta}) p(v_{\theta}) d v_{\theta}. \label{eq:z_dot}
\end{equation}

As shown in \cite{montbrio_macroscopic_2015}, this integral can be evaluated analytically if $p(v_{\theta})$ is chosen as a Lorentzian density function

\begin{equation}
    p(v_{\theta}) = \frac{1}{\pi} \frac{\Delta_v}{[v_{\theta} - \bar v_{\theta}]^2 + \Delta_v^2}, \label{eq:p_lorentzian}
\end{equation}

centered at $\bar v_{\theta}$ and with half-width-at-half-maximum $\Delta_v$.
For this particular choice, we can solve eq.\eqref{eq:z_dot} by evaluating $\frac{\partial}{\partial t} z(t,v_{\theta})$ at $v_{\theta} = \bar v_{\theta} + i \Delta_v$.
Further consideration of eqs.\eqref{eq:rx} and \eqref{eq:vy}, yields $\frac{\pi C}{k} r(t) + iv(t) = z(t,\bar v_{\theta} + i \Delta_v)$, leading to the following set of coupled ordinary differential equations 

\begin{align}
    C \dot{r} = &\frac{\Delta_v k^2}{\pi C} (v-v_r) + r(k(2 v - v_r - \bar v_{\theta}) - g s), \label{eq:r_dot}\\
    C \dot{v} = &k v(v -v_r-\bar v_{\theta}) - \pi C r(\Delta_v + \frac{\pi C}{k} r) \label{eq:v_dot}\\
    &+ v_r \bar v_{\theta} - u + I  + g s (E-v), \nonumber \\
    \tau_u \dot{u} = &b(v-v_r) - u + \tau_u \kappa r, \label{eq:u_dot}\\
    \tau_s \dot{s} = &-s + \tau_s J r, \label{eq:s_dot}
\end{align}

which fully captures the mean-field dynamics of the IK neuron network given by eqs.(\ref{eq:v_i}-\ref{eq:s}).
In the derivation of eqs.\eqref{eq:s_dot} and \eqref{eq:u_dot}, we used $r(t) = \frac{1}{N} \sum_{j=1}^{N} \delta(v_j(t)-v_p)$ and $v(t) = \frac{1}{N}\sum_{j=1}^N v_j(t)$.

For comparison, we also consider heterogeneity in the input parameter $I_i = \eta_i + I_{ext}$. 
Under the assumption that eq.\eqref{eq:z} depends on the quenched neuron-specific parameter $\eta$ instead of $v_{\theta}$, and assuming a Lorentzian density $p(\eta)$
\begin{equation}
    p(\eta) = \frac{1}{\pi} \frac{\Delta_{\eta}}{[\eta - \bar \eta]^2 + \Delta_{\eta}^2}, \label{eq:eta_lorentzian}
\end{equation}

we can integrate eq.\eqref{eq:z} with respect to $p(\eta)$ and obtain the following set of mean-field equations

\begin{align}
    C \dot r = &\frac{\Delta_{\eta} k^2}{\pi C} + r(k(2 v - v_r - \bar v_{\theta}) - g s), \label{eq:r2_dot}\\
   C \dot{v} = &k v (v -v_r-\bar v_{\theta}) - \frac{(\pi C r)^2}{k} \label{eq:v2_dot}\\
    &+ v_r \bar v_{\theta} - u + I  + g s (E-v), \nonumber \\
    \tau_u \dot{u} = &b(v-v_r) -u + \tau_u \kappa r, \label{eq:u2_dot}\\
    \tau_s \dot s = &-s + \tau_s J r. \label{eq:s2_dot}
\end{align}

\section{Effects of Neural Heterogeneity on Neural Population Dynamics}

To study the effects of distributed firing thresholds $v_{\theta}$ on the dynamics of a population of coupled IK neurons via the mean-field equations (\ref{eq:r_dot}-\ref{eq:s_dot}), we analyze how the parameter $\Delta_v$ affects the system dynamics.
This parameter represents the width of the distribution of $v_{\theta}$ and it thus controls the amount of heterogeneity within the population.

As a first result, we find that $\Delta_v$ directly affects the dynamics of both $r$ and $v$.
Additionally, the impact of $\Delta_v$ on the system is state-dependent, since it scales with the values of both $v$ and $r$.
This result is an important difference between our mean-field model and previously derived mean-field models of spiking neural networks, such as \cite{montbrio_macroscopic_2015,gast_mean-field_2020}. 
These previous models are similar in form to eqs.(\ref{eq:r2_dot}-\ref{eq:s2_dot}), where $\Delta_{\eta}$ only affects the dynamics in $r$ and does so in a state-independent way, i.e. it acts as a mere offset to $\dot r$ and does not scale with any state variable.
In this previously investigated class of models, changes in neural heterogeneity merely lead to a shift in the average firing rate of the system.
In contrast, the effect of heterogeneity in the macroscopic dynamics of our model depends on the dynamic regime of the population. 
When the population membrane potential $v$ is close to its resting value $v_r$, $\Delta_v$ will barely affect the population firing rate, since $v-v_r$ will vanish and $r$ will be small.
However, when the population is in a de- or hyper-polarized state, the impact of $\Delta_v$ is more complicated to predict, since it affects eqs.\eqref{eq:r_dot} and \eqref{eq:v_dot} in different ways.
For example, if the system is in a regime of strong depolarization, $\Delta_v$ has a positive effect on the firing rate but a hyper-polarizing effect on the membrane potential.

To improve our understanding of this source of neural heterogeneity on the population dynamics, we studied its effects via bifurcation analysis using the dynamical systems modeling software PyRates \cite{gast_pyratespython_2019}.
For each type of model cell, we selected parameters similar to those fit in \cite{izhikevich_dynamical_2007}; the full set of parameters used for each neuron type can be found in Tabs.\ref{tab:rs}-\ref{tab:lts}.

\subsection{Mean-Field Dynamics of Heterogeneous Regular-Spiking Neurons}

\begin{figure}
    \centering
    \includegraphics[width=1.0\columnwidth]{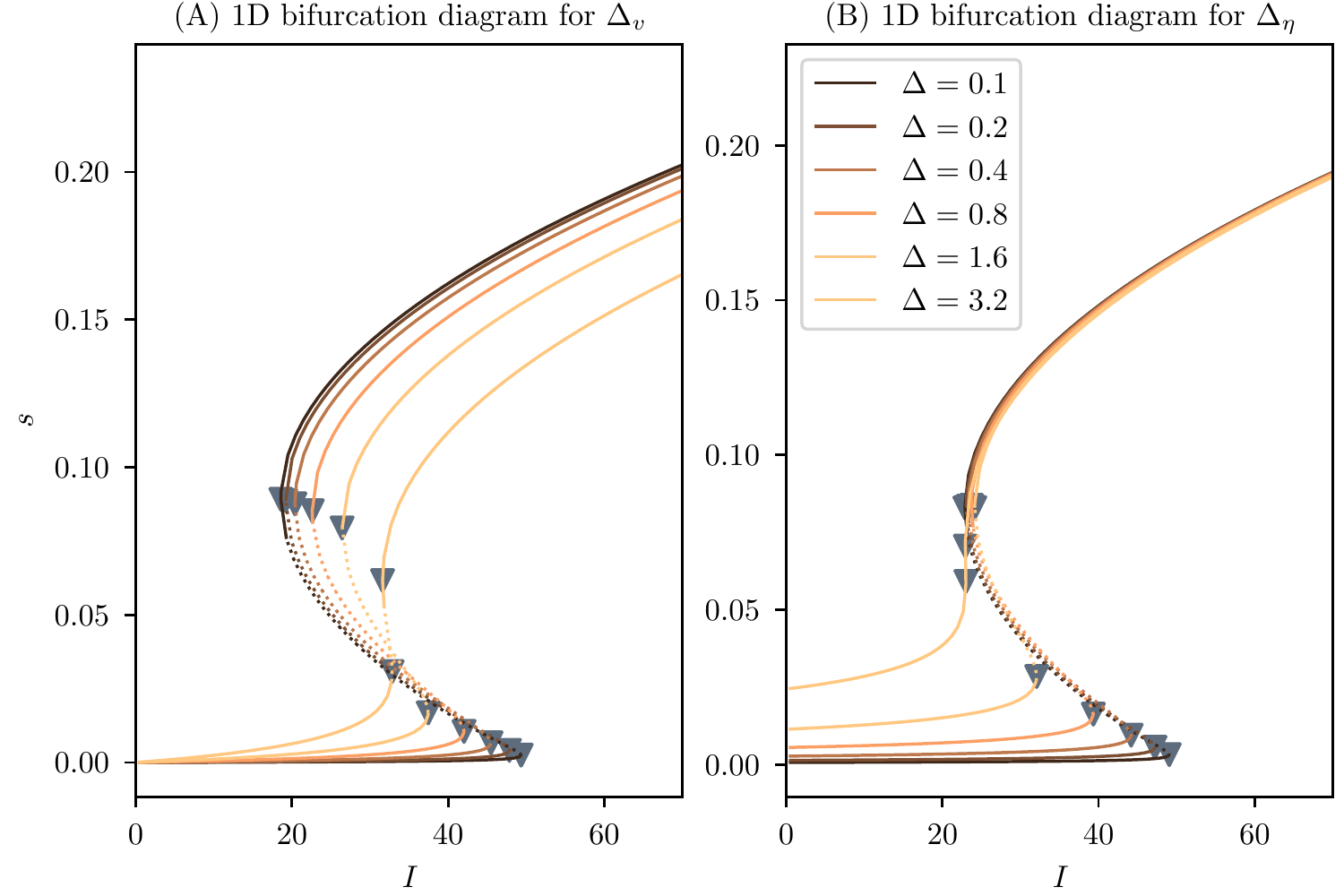}
    \caption{\textit{Effects of different forms of neural heterogeneity on a population of regular-spiking neurons.} \textbf{(A)} 1D bifurcation diagram for the state variable $s$ of the mean-field model given by eqs.(\ref{eq:r_dot}-\ref{eq:s_dot}) as a function of the bifurcation parameter $I$. Solid (dotted) lines represent stable (unstable) steady-state solutions. Grey triangles mark fold bifurcation points. Color coded curves represent different degrees of heterogeneity. \textbf{(B)} Same as (A) but for the mean-field model given by eqs.(\ref{eq:r2_dot}-\ref{eq:s2_dot}).}
    \label{fig:rs}
\end{figure}

We first investigated the effect of "within-type" heterogeneity on the dynamics of a population of excitatory regular-spiking neurons.
As shown in Fig.\ref{fig:rs}, populations of regular-spiking neurons exhibit fold bifurcations that give rise to a bi-stable regime where changes in the background current $I$ switch the population between stable down- and up-states.
Increasing the degree of neural heterogeneity linearizes the response of the system to changes in $I$, thus causing the bi-stable regime to shrink and eventually collapse in a cusp bifurcation.
This reduction of the range of values of $I$ for which the bi-stable regime exists, occurs for increases in either $\Delta_v$ or $\Delta_{\eta}$; however, only the former affects the firing rate of the system in the up-state (Fig.\ref{fig:rs}A-B).
The general bifurcation structure of the system does not differ between the model with spike-threshold heterogeneity (eqs.(\ref{eq:r_dot}-\ref{eq:s_dot})) and the model with input heterogeneity (eqs.(\ref{eq:r2_dot}-\ref{eq:s2_dot})); 
this holds true not only about bi-stability but also for other dynamic regimes of the model.
Given this observation, and given that spike-threshold heterogeneity (as opposed to background current heterogeneity) can be directly measured across single neurons, we restrict the results reported below to the IK model with spike-threshold heterogeneity (eqs.(\ref{eq:r_dot}-\ref{eq:s_dot})), which allows for a better comparison to experimental data.

\subsection{Mean-Field Dynamics of Heterogeneous Fast-Spiking Neurons}

We next investigated the effect of "within-type" heterogeneity on the dynamics of a population of inhibitory neurons, using a model of fast-spiking inhibitory interneurons\cite{izhikevich_dynamical_2007}.
Recurrent inhibitory populations are known to exhibit synchronized oscillations given sufficient excitatory drive \cite{golomb_dynamics_1993,gast_role_2021}; here we examined how the transition from asynchronous dynamics to synchronous oscillations is affected by spike-threshold heterogeneity ($\Delta_v$).

\begin{figure}
    \centering
    \includegraphics[width=1.0\columnwidth]{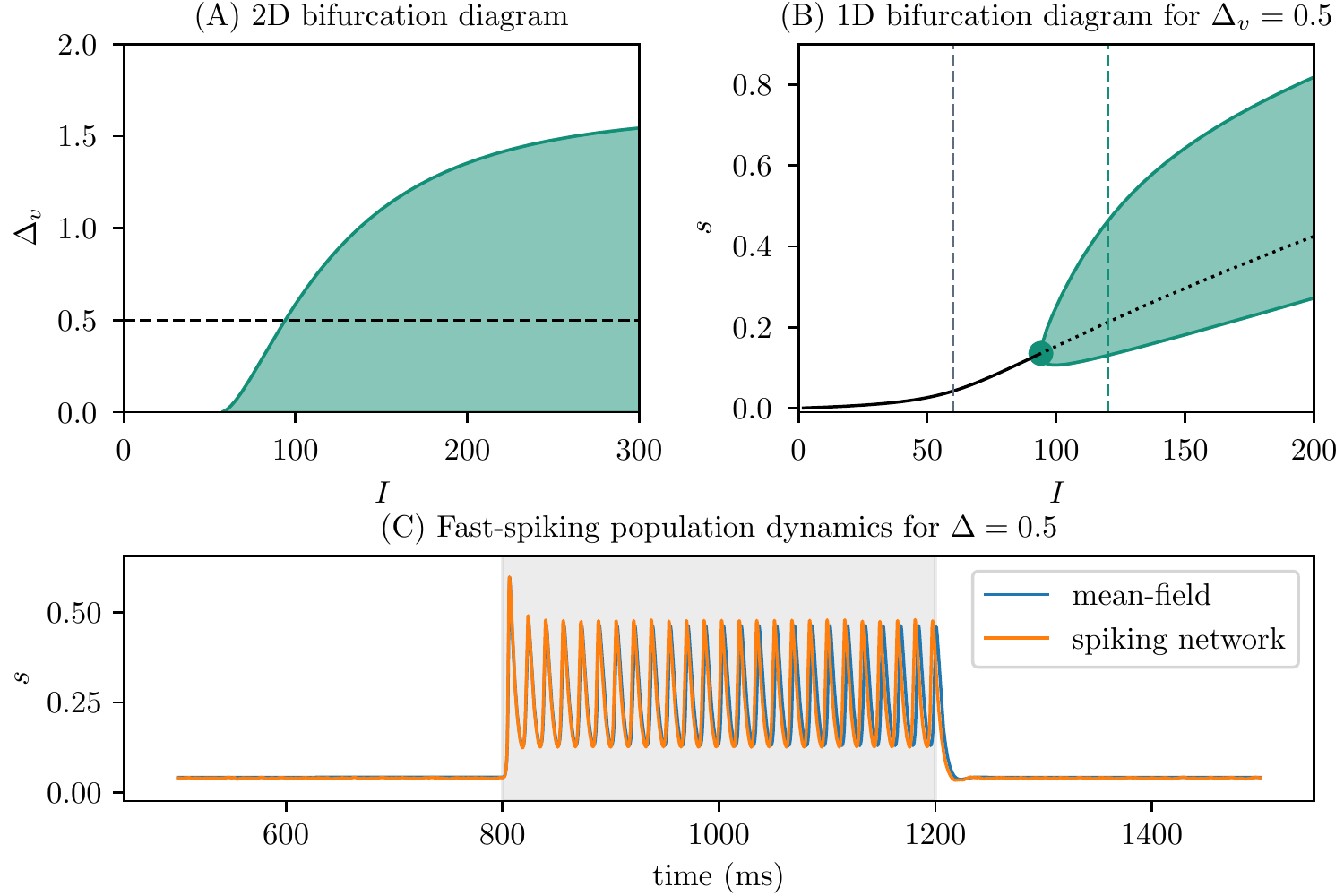}
    \caption{\textit{Desynchronization of fast-spiking neurons by increased spike-threshold heterogeneity.} \textbf{(A)} 2D bifurcation diagram for the heterogeneity parameter $\Delta_v$ and the global input $I$. The regions filled in white and green mark the asynchronous and synchronous-oscillatory regimes, respectively. The dashed line represents the 1D cut depicted in B. \textbf{(B)} 1D bifurcation diagram for the state variable $s$ of the mean-field model given by eqs.(\ref{eq:r_dot}-\ref{eq:s_dot}) as a function of the bifurcation parameter $I$. Solid (dotted) lines represent stable (unstable) steady-state solutions. The green region depicts limit cycle amplitudes of the oscillatory regime emerging from the supercritical Andronov-Hopf bifurcation at the green circle. The dashed lines represent the two dynamic regimes depicted in C. \textbf{(C)} Simulated responses of the mean-field model and corresponding spiking neural network as the input is stepped from $I = 60$ to $I = 120$ at $t= 800$ and back at $t= 1200$ (gray shaded region).}
    \label{fig:fs}
\end{figure}

Fig.\ref{fig:fs}A and B show a supercritical Andronov-Hopf bifurcation that gives rise to a synchronous-oscillatory regime as excitatory input to the system increases.
The strength of the excitatory input $I$ required to push the system into the oscillatory regime increases with spike-threshold heterogeneity, until the oscillatory regime eventually vanishes.

Fig.\ref{fig:fs}C illustrates the good agreement between the spiking neural network dynamics and the dynamics of the mean-field model; the transition between asynchronous and synchronous-oscillatory regimes occurs as predicted by the bifurcation analysis.
This agreement is obtained in spite of the fact by incorporating biophysically derived model parameters we break some of the assumptions of the original mean-field model derivation, namely that neuron spiking and reset values are at $\pm\infty$.

\subsection{Effects of Neural Heterogeneity on Excitatory-Inhibitory Circuits}

In many brain regions, populations of excitatory and inhibitory neurons interact within local microcircuits \cite{harris_neocortical_2015}.
Such microcircuits have been proposed to play a crucial role in the integration of bottom-up and top-down information and are at the heart of many theories of brain function \cite{bastos_canonical_2012,wolf_dynamical_2014,vogels_gating_2009,schmidt_network_2018,taher_exact_2020}.
Our next analysis concerns the dynamics of systems of excitatory-inhibitory circuits and how between-type and within-type heterogeneity interact in such circuits.

\subsubsection{Coupled Regular- and Fast-Spiking Populations}

\begin{figure*}
    \centering
    \includegraphics[width=1.0\textwidth]{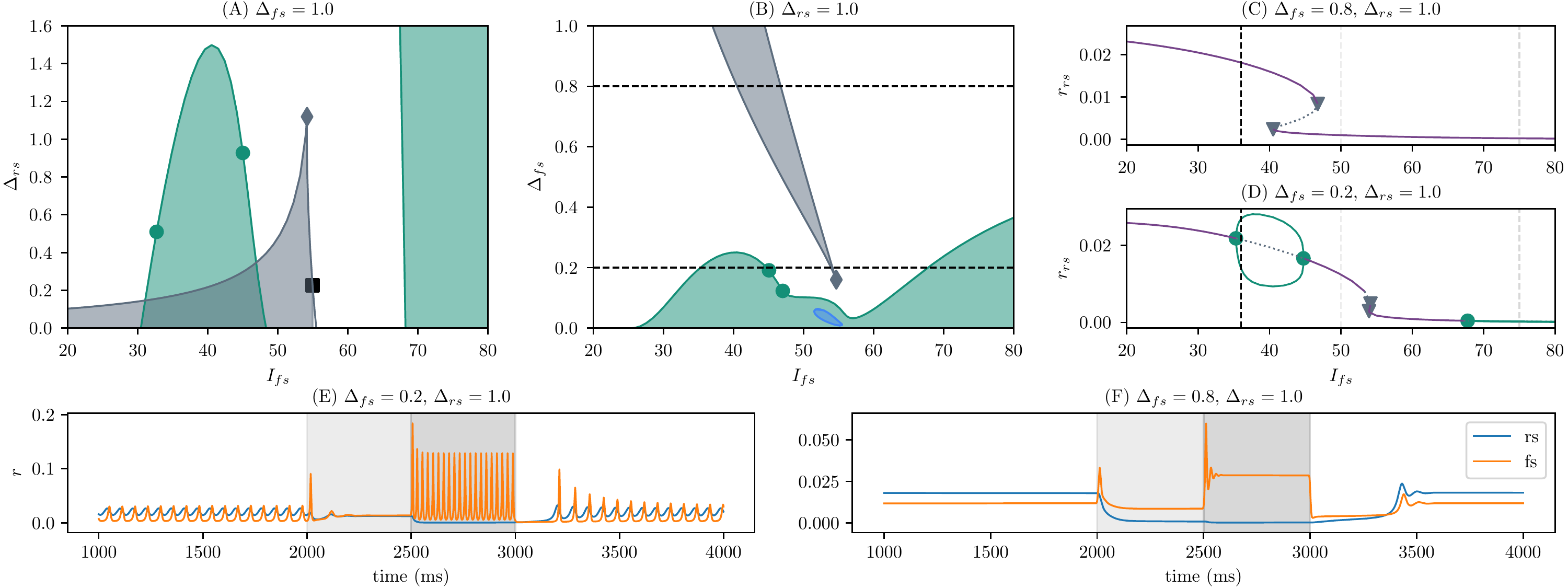}
    \caption{\textit{Fast-spiking heterogeneity separates bi-stable from synchronous oscillatory regimes in coupled excitatory-inhibitory networks.} \textbf{(A, B)} 2D bifurcation diagrams. Parameter space regions filled in grey and green depict bi-stable and synchronous oscillatory regimes, respectively. Regions filled in blue depict regions where two synchronous oscillatory regimes collide in a period doubling bifurcation. Grey diamonds represent cusp bifurcations, the black square represents a Bogdanov-Takens bifurcation, and green circles represent generalized Hopf bifurcations. The parameters correspond to fast-spiking ($fs$) or regular-spiking ($rs$) subpopulations. \textbf{(C, D)} 1D bifurcation diagrams for two different levels of the threshold heterogeneity of fast-spiking neurons. \textbf{(E, F)} System dynamics for two different levels of fast-spiking heterogeneity in response to changes in the input to fast-spiking neurons. Inputs were chosen as $I_{rs} = 50 \text{ } \forall t$, $I_{fs} = 50$ for $2000 < t < 2500$, $I_{fs} = 75$ for $2500 < t < 3000$, and $I_{fs} = 36$ otherwise.}
    \label{fig:eic}
\end{figure*}

We first studied a system of interacting regular-spiking excitatory and fast-spiking inhibitory neurons. 
Each population was governed by the IK mean-field eqs.(\ref{eq:r_dot}-\ref{eq:s_dot}).
The full set of equations for the mean-field model and the corresponding spiking neural network are given in Appendix A, eqs.\eqref{eq:eic_mf} and eqs.\eqref{eq:eic_snn}, respectively; the synaptic coupling parameters are reported in Tab.\ref{tab:two_pop_coupling}.
As shown in Fig.~\ref{fig:eic}, this system exhibits asynchronous, synchronous-oscillatory, and bi-stable regimes, as characteristic of the independent dynamics of regular-spiking and fast-spiking populations.
Because regular-spiking and fast-spiking neuron within-type heterogeneities have distinct effects on the transitions between these regimes, we find that between- and within type heterogeneity interact in this population composed of two types of neurons.
As discussed above, heterogeneity within the regular-spiking neurons decreases the extent of the bi-stable regime, whereas heterogeneity within the fast-spiking neurons reduces the prevalence of the synchronous-oscillatory regime (Fig.~\ref{fig:eic}A vs. B). 
Interestingly, heterogeneity within fast-spiking inhibitory neurons separates the synchronous-oscillatory regime from the bi-stable regime in the parameter plane spanned by fast-spiking parameters $\Delta_{fs}$ and $I_{fs}$ (green versus grey regions in Fig.~\ref{fig:eic}B).
The heterogeneity in fast-spiking interneurons thus determines whether they can force the regular-spiking population to a down-state (Fig.~\ref{fig:eic}C and F), or whether they induce high-frequency oscillations in the system (Fig.~\ref{fig:eic}D and E).

\subsubsection{Coupled Excitatory Regular-Spiking, Inhibitory Fast-Spiking, and Low-Threshold-Spiking Populations}

To examine whether our results regarding the control effects of fast-spiking heterogeneity are specific to that particular interneuron type or can be generalized to other types of inhibitory interneuron populations, we consider a three-population model containing regular-spiking, fast-spiking, and low-threshold spiking neurons.
Low-threshold spiking neurons differ substantially in their electrophysiological properties from fast-spiking neurons \cite{izhikevich_dynamical_2007,harris_neocortical_2015} and result in an additional time scale for inhibition. 
Experimental studies suggest that fast-spiking and low-threshold spiking neurons have distinct effects on neural network computations and oscillations, and thus affect behavior differentially \cite{wilson_division_2012,kvitsiani_distinct_2013,chen_distinct_2017}.

Here, we examine how the within-type heterogeneities of these distinct inhibitory interneuron types interact to shape mesoscopic circuit dynamics.
To this end, we examined the effects of the low-threshold spiking population on the circuit dynamics for two different levels of fast-spiking neuron threshold heterogeneity: $\Delta_{fs}$ = 0.4 and $\Delta_{fs}$ = 0.8.
The mean-field and spiking network model equations are given in Appendix B by eqs.\ref{eq:eiic_mf} and eqs.\ref{eq:eiic_snn}, respectively, with synaptic coupling parameters as reported in Tab.\ref{tab:three_pop_coupling} (Appendix C).

\begin{figure*}
    \centering
    \includegraphics[width=1.0\textwidth]{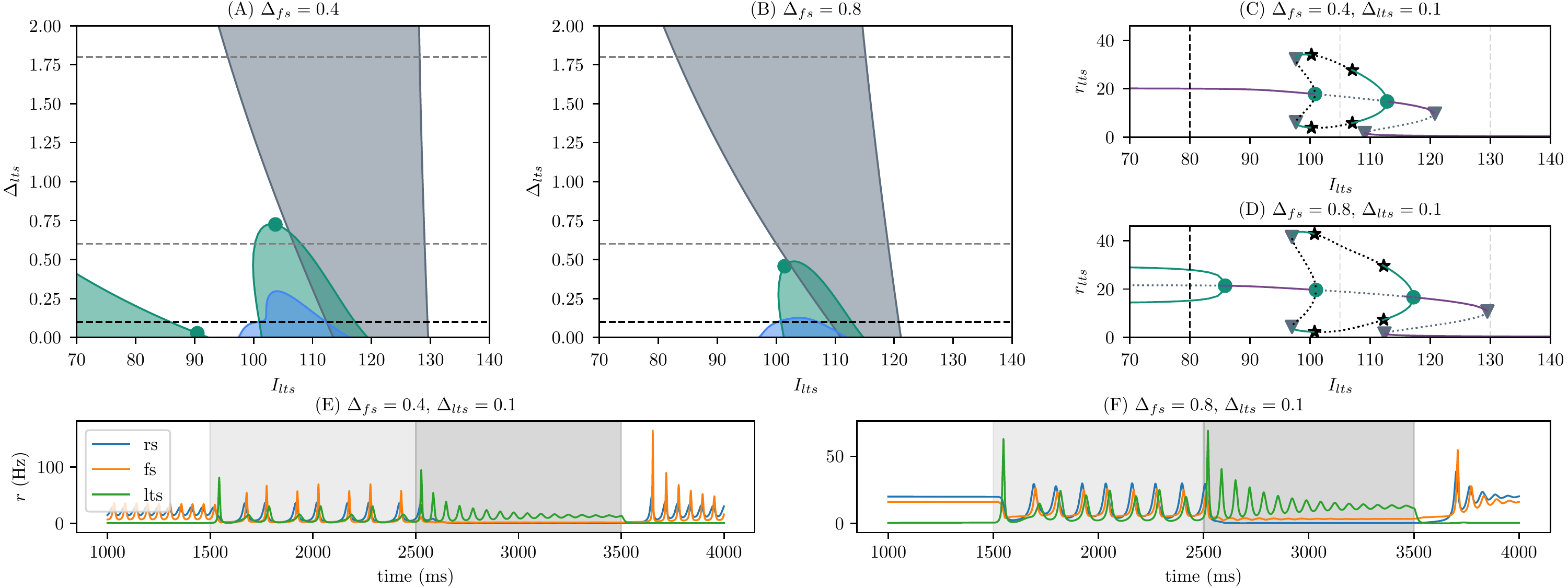}
    \caption{\textit{Heterogeneity of inhibitory interneuron populations controls mesoscopic excitatory-inhibitory circuit regimes.} \textbf{(A, B)} 2D bifurcation diagrams use the same color and symbol conventions as in Fig.\ref{fig:eic} A and B. Horizontal dashed lines represent values of $\Delta_{lts}$ that correspond to spike threshold variances reported in different cortical layers in rats \cite{wang_anatomical_2004}. \textbf{(C, D)} 1D bifurcation diagrams for two different levels of fast-spiking heterogeneity at a low degree of low-threshold spiking heterogeneity.
    Black stars depict period doubling bifurcations, whereas the remaining symbols follow the same conventions as in Fig.\ref{fig:eic}.
    \textbf{(E, F)} System dynamics for two different levels of fast-spiking heterogeneity in response to changes in the input to the low-threshold spiking population. Inputs were chosen as $I_{rs} = 60 \text{ } \forall t$, $I_{fs} = 40 \text{ } \forall t$, $I_{lts} = 105$ for $1500 < t < 2500$, $I_{lts} = 130$ for $2500 < t < 3500$, and $I_{lts} = 80$ otherwise.}
    \label{fig:eiic}
\end{figure*}

The width of the spike threshold distributions for all three model populations were taken from spike threshold variances reported in the literature; therefore, our findings reflect dynamic regimes expected to be found in cortical microcircuits. 
Specifically, the two values of $\Delta_{fs}$ used for Fig.\ref{fig:eiic} are fitted to sample variances reported for in-vitro recordings of fast-spiking neurons in deep vs. superficial cortical layers in mice \cite{lau_impaired_2000}.
The dashed horizontal lines in Fig.~\ref{fig:eiic}A and B represent values of $\Delta_{lts}$ that correspond to the sample variances reported for in-vitro recordings from low-threshold spiking neurons in different layers of somatosensory cortex in rats \cite{wang_anatomical_2004}.

As shown in Fig.~\ref{fig:eiic}A and B, we find that threshold heterogeneity in both fast-spiking and low-threshold spiking interneurons has a de-synchronizing effect on the network dynamics.
Depending on whether $\Delta_{fs} = 0.4$ or $\Delta_{fs} = 0.8$, two or one synchronous oscillatory regimes exist in the parameter space examined in Fig.~\ref{fig:eiic}.
One oscillatory regime emerges from the synaptic interactions between regular-spiking and fast-spiking neurons, whereas the other emerges from the interaction between all three populations (Fig.~\ref{fig:eiic}D and E).
For these synchronous oscillatory regimes to be accessed via extrinsic inputs to the low-threshold spiking neuron population, both low-threshold and fast-spiking neuron heterogeneity must be sufficiently small.
The two synchronous oscillatory regimes can also collide to form a period doubling bifurcation (see Fig.~\ref{fig:eiic}A-D) in a small region of the parameter space spanned by $\Delta_{lts}$ and $I_{lts}$.
A comparison between Fig.~\ref{fig:eiic}E and F, reveals that the spectral properties of the emerging limit cycle critically depend on the degree of neural heterogeneity.
Note also that the within-type heterogeneity in the three model populations can have opposing effects on the width of the bi-stable regime:
whereas increased regular spiking heterogeneity always leads to a reduced range of the bi-stable regime (see Fig.~\ref{fig:rs} and Fig.~\ref{fig:eic}A), increased fast-spiking and low-threshold spiking neuron heterogeneity can lead to increased range of the bi-stable regime (see Fig.~\ref{fig:eic}B and Fig.~\ref{fig:eiic}A and B).

Together, our results illustrate that spike threshold heterogeneities within each of the different inhibitory interneuron populations interact in complex ways to shape the dynamic landscape of mesoscopic circuits. 

\section{Conclusion}

The dynamics of spiking neural networks are shaped by both within- and between-type neural heterogeneity.
In this study we showed that the effects of within-type heterogeneity on network dynamics strongly depend on the type of neuron affected.
The heterogeneity of excitatory neurons affects the co-existence of and phase transitions between high- and low-activity network activity in bi-stable regimes; in contrast, the heterogeneity of inhibitory populations most strongly affects the synchronization properties of the network. 

Our study of coupled excitatory and inhibitory populations further suggests that the de-synchronizing effects of inhibitory population heterogeneity can play a critical role at the level of mesoscopic brain circuits.
In coupled excitatory-inhibitory networks, the degree of heterogeneity in the inhibitory interneuron populations determines whether the circuit exhibits one of two dynamical regimes. 
When heterogeneity is low, the system falls into synchronized oscillations at characteristic frequencies, a property that might facilitate information gating by interneuron populations \cite{amilhon_parvalbumin_2015,veit_cortical_2017,hahn_portraits_2019}. 
When heterogeneity is high, oscillations are eliminated and the network can quickly switch between high- and low-activity states; the ability of neural populations to perform such rapid transitions is important in working memory models \cite{schmidt_network_2018,taher_exact_2020}.
Our findings therefore show the importance of experimentally quantifying neuronal heterogeneity within recorded populations, and suggest that including these measures could help account for diverse population-specific dynamics across brain regions.
The model derived in this letter is well suited to explain and interpret neural recordings because it is built upon biophysical variables whose model values can be directly informed by such recordings.

\bibliographystyle{apsrev4-1}

\bibliography{references}

\newpage
\appendix

\section{Appendixes}

\subsection{Appendix A: Two population model equations}

Here we present the full set of equations used for the two-population EIC model.
The SNN dynamics of a system of all-to-all coupled excitatory regular-spiking and inhibitory fast-spiking neurons are given by:

\begin{subequations}
\begin{align}
     C^{\text{rs}} \frac{d}{dt} v_i^{\text{rs}} = &k^{\text{rs}} (v_i^{\text{rs}} - v_r^{\text{rs}})(v_i^{\text{rs}} - v_{\theta}^{\text{rs}}) - u^{\text{rs}} + I^{\text{rs}}\\
     &+ J_{\text{rr}} g^{\text{rs}} s^{\text{rs}} (E^{\text{rs}}-v_i^{\text{rs}}) + J_{\text{rf}} g^{\text{fs}} s^{\text{fs}} (E^{\text{fs}}-v_i^{\text{rs}}), \nonumber\\
    \tau_u^{\text{rs}} \frac{d}{dt} u^{\text{rs}} = &b^{\text{rs}}(\frac{1}{N}\sum_{j=1}^N v_j^{\text{rs}} - v_r^{\text{rs}}) - u^{\text{rs}}\\
    &+ \frac{\tau_u^{\text{rs}} \kappa^{\text{rs}}}{N} \sum_{j=1}^N \delta(v_j^{\text{rs}}-v_p^{\text{rs}}), \nonumber\\
    \tau_s^{\text{rs}} \frac{d}{dt} s^{\text{rs}} = &-s^{\text{rs}} + \frac{\tau_s^{\text{rs}}}{N} \sum_{j=1}^{N} \delta(v_j^{\text{rs}}-v_p^{\text{rs}}),\\
    C^{\text{fs}} \frac{d}{dt} v_i^{\text{fs}} = &k^{\text{fs}} (v_i^{\text{fs}} - v_r^{\text{fs}})(v_i^{\text{fs}} - v_{\theta}^{\text{fs}}) - u^{\text{fs}} + I^{\text{fs}}\\
     &+ J_{\text{fr}} g^{\text{rs}} s^{\text{rs}} (E^{\text{rs}}-v_i^{\text{fs}}) + J_{\text{ff}} g^{\text{fs}} s^{\text{fs}} (E^{\text{fs}}-v_i^{\text{fs}}), \nonumber\\
    \tau_u^{\text{fs}} \frac{d}{dt} u^{\text{fs}} = &b^{\text{fs}}(\frac{1}{N}\sum_{j=1}^N v_j^{\text{fs}} - v_r^{\text{fs}}) - u^{\text{fs}}\\
    &+ \frac{\tau_u^{\text{fs}} \kappa^{\text{fs}}}{N} \sum_{j=1}^N \delta(v_j^{\text{fs}}-v_p^{\text{fs}}), \nonumber\\
    \tau_s^{\text{fs}} \frac{d}{dt} s^{\text{fs}} = &-s^{\text{fs}} + \frac{\tau_s^{\text{fs}}}{N} \sum_{j=1}^{N} \delta(v_j^{\text{fs}}-v_p^{\text{fs}}),
\end{align}
\label{eq:eic_snn}
\end{subequations}

where the superscripts $\text{rs}$ and $\text{fs}$ denote regular-spiking and fast-spiking populations, respectively.
The mean-field equations for this SNN can be derived following the procedure outlined in the main text, to obtain: 

\begin{subequations}
\begin{align}
    C^{\text{rs}} \frac{d}{dt}r^{\text{rs}} = &\frac{\Delta_v^{\text{rs}} (k^{\text{rs}})^2}{\pi C^{\text{rs}}} (v^{\text{rs}}-v_r^{\text{rs}})\\ 
    &+ r^{\text{rs}}(k^{\text{rs}}(2 v^{\text{rs}} - v_r^{\text{rs}} - \bar{v_{\theta}^{\text{rs}}}) - J_{\text{rr}} g^{\text{rs}} s^{\text{rs}} - J_{\text{rf}} g^{\text{fs}} s^{\text{fs}}), \nonumber\\
    C^{\text{rs}} \frac{d}{dt} v^{\text{rs}} = &k^{\text{rs}} v^{\text{rs}}(v^{\text{rs}} -v_r^{\text{rs}}-\bar{v_{\theta}^{\text{rs}}})\\ 
    &- \pi C^{\text{rs}} r^{\text{rs}}(\Delta_v^{\text{rs}} + \frac{\pi C^{\text{rs}}}{k^{\text{rs}}} r^{\text{rs}}) + v_r^{\text{rs}} \bar{v_{\theta}^{\text{rs}}} - u^{\text{rs}} + I^{\text{rs}} \nonumber\\
    &+ J_{\text{rr}} g^{\text{rs}} s^{\text{rs}} (E^{\text{rs}}-v^{\text{rs}}) + J_{\text{rf}} g^{\text{fs}} s^{\text{fs}} (E^{\text{fs}}-v^{\text{rs}}), \nonumber\\
    \tau_u^{\text{rs}} \frac{d}{dt}u^{\text{rs}} = &b^{\text{rs}}(v^{\text{rs}}-v_r^{\text{rs}}) - u^{\text{rs}} + \tau_u^{\text{rs}} \kappa^{\text{rs}} r^{\text{rs}},\\
    \tau_s^{\text{rs}} \frac{d}{dt} s^{\text{rs}} = &-s^{\text{rs}} + \tau_s^{\text{rs}} r^{\text{rs}},\\
    C^{\text{fs}} \frac{d}{dt}r^{\text{fs}} = &\frac{\Delta_v^{\text{fs}} (k^{\text{fs}})^2}{\pi C^{\text{fs}}} (v^{\text{fs}}-v_r^{\text{fs}})\\ 
    &+ r^{\text{fs}}(k^{\text{fs}}(2 v^{\text{fs}} - v_r^{\text{fs}} - \bar{v_{\theta}^{\text{fs}}}) - J_{\text{fr}} g^{\text{rs}} s^{\text{rs}} - J_{\text{ff}} g^{\text{fs}} s^{\text{fs}}), \nonumber\\
    C^{\text{fs}} \frac{d}{dt} v^{\text{fs}} = &k^{\text{fs}} v^{\text{fs}}(v^{\text{fs}} -v_r^{\text{fs}}-\bar{v_{\theta}^{\text{fs}}})\\ 
    &- \pi C^{\text{fs}} r^{\text{fs}}(\Delta_v^{\text{fs}} + \frac{\pi C^{\text{fs}}}{k^{\text{fs}}} r^{\text{fs}}) + v_r^{\text{fs}} \bar{v_{\theta}^{\text{fs}}} - u^{\text{fs}} + I^{\text{fs}} \nonumber\\
    &+ J_{\text{fr}} g^{\text{rs}} s^{\text{rs}} (E^{\text{rs}}-v^{\text{fs}}) + J_{\text{ff}} g^{\text{fs}} s^{\text{fs}} (E^{\text{fs}}-v^{\text{fs}}), \nonumber\\
    \tau_u^{\text{fs}} \frac{d}{dt}u^{\text{fs}} = &b^{\text{fs}}(v^{\text{fs}}-v_r^{\text{fs}}) - u^{\text{fs}} + \tau_u^{\text{fs}} \kappa^{\text{fs}} r^{\text{fs}},\\
    \tau_s^{\text{fs}} \frac{d}{dt} s^{\text{fs}} = &-s^{\text{fs}} + \tau_s^{\text{fs}} r^{\text{fs}}.
\end{align}
\label{eq:eic_mf}
\end{subequations}

\subsection{Appendix B: Three population model equations\label{app:eiic}}

Here we present the full set of equations used for the three-population EIC model.
The SNN dynamics of a system of all-to-all coupled excitatory regular-spiking, inhibitory fast-spiking, and inhibitory low-threshold spiking neurons are given by:

\begin{subequations}
\begin{align}
     C^{\text{rs}} \frac{d}{dt} v_i^{\text{rs}} = &k^{\text{rs}} (v_i^{\text{rs}} - v_r^{\text{rs}})(v_i^{\text{rs}} - v_{\theta}^{\text{rs}}) - u^{\text{rs}} + I^{\text{rs}}\\ 
     &+ J_{\text{rr}} g^{\text{rs}} s^{\text{rs}} (E^{\text{rs}}-v_i^{\text{rs}}) + J_{\text{rf}} g^{\text{fs}} s^{\text{fs}} (E^{\text{fs}}-v_i^{\text{rs}}) \nonumber\\
     &+ J_{\text{rl}} g^{\text{lts}} s^{\text{lts}} (E^{\text{lts}}-v_i^{\text{rs}}), \nonumber\\
    \tau_u^{\text{rs}} \frac{d}{dt} u^{\text{rs}} = &b^{\text{rs}}(\frac{1}{N}\sum_{j=1}^N v_j^{\text{rs}} - v_r^{\text{rs}}) - u^{\text{rs}}\\
    &+ \frac{\tau_u^{\text{rs}} \kappa^{\text{rs}}}{N} \sum_{j=1}^N \delta(v_j^{\text{rs}}-v_p^{\text{rs}}), \nonumber\\
    \tau_s^{\text{rs}} \frac{d}{dt} s^{\text{rs}} = &-s^{\text{rs}} + \frac{\tau_s^{\text{rs}}}{N} \sum_{j=1}^{N} \delta(v_j^{\text{rs}}-v_p^{\text{rs}}),\\
    C^{\text{fs}} \frac{d}{dt} v_i^{\text{fs}} = &k^{\text{fs}} (v_i^{\text{fs}} - v_r^{\text{fs}})(v_i^{\text{fs}} - v_{\theta}^{\text{fs}}) - u^{\text{fs}} + I^{\text{fs}}\\
     &+ J_{\text{fr}} g^{\text{rs}} s^{\text{rs}} (E^{\text{rs}}-v_i^{\text{fs}}) + J_{\text{ff}} g^{\text{fs}} s^{\text{fs}} (E^{\text{fs}}-v_i^{\text{fs}}) \nonumber\\
     &+ J_{\text{fl}} g^{\text{lts}} s^{\text{lts}} (E^{\text{lts}}-v_i^{\text{fs}}), \nonumber\\
    \tau_u^{\text{fs}} \frac{d}{dt} u^{\text{fs}} = &b^{\text{fs}}(\frac{1}{N}\sum_{j=1}^N v_j^{\text{fs}} - v_r^{\text{fs}}) - u^{\text{fs}}\\
    &+ \frac{\tau_u^{\text{fs}} \kappa^{\text{fs}}}{N} \sum_{j=1}^N \delta(v_j^{\text{fs}}-v_p^{\text{fs}}), \nonumber\\
    \tau_s^{\text{fs}} \frac{d}{dt} s^{\text{fs}} = &-s^{\text{fs}} + \frac{\tau_s^{\text{fs}}}{N} \sum_{j=1}^{N} \delta(v_j^{\text{fs}}-v_p^{\text{fs}}),\\
    C^{\text{lts}} \frac{d}{dt} v_i^{\text{lts}} = &k^{\text{lts}} (v_i^{\text{lts}} - v_r^{\text{lts}})(v_i^{\text{lts}} - v_{\theta}^{\text{lts}}) - u^{\text{lts}} + I^{\text{lts}}\\
     &+ J_{\text{lr}} g^{\text{rs}} s^{\text{rs}} (E^{\text{rs}}-v_i^{\text{lts}}) + J_{\text{lf}} g^{\text{fs}} s^{\text{fs}} (E^{\text{fs}}-v_i^{\text{lts}}) \nonumber\\
    \tau_u^{\text{lts}} \frac{d}{dt} u^{\text{lts}} = &b^{\text{lts}}(\frac{1}{N}\sum_{j=1}^N v_j^{\text{lts}} - v_r^{\text{lts}}) - u^{\text{lts}}\\
    &+ \frac{\tau_u^{\text{lts}} \kappa^{\text{lts}}}{N} \sum_{j=1}^N \delta(v_j^{\text{lts}}-v_p^{\text{lts}}), \nonumber\\
    \tau_s^{\text{lts}} \frac{d}{dt} s^{\text{lts}} = &-s^{\text{lts}} + \frac{\tau_s^{\text{lts}}}{N} \sum_{j=1}^{N} \delta(v_j^{\text{lts}}-v_p^{\text{lts}}),
\end{align}
\label{eq:eiic_snn}
\end{subequations}

where the superscripts $\text{rs}$, $\text{fs}$, and $\text{lts}$ denote regular-spiking, fast-spiking, and low-threshold spiking populations, respectively.
The mean-field equations for this SNN can be derived following the procedure outlined in the main text, to obtain: 

\begin{subequations}
\begin{align}
    C^{\text{rs}} \frac{d}{dt}r^{\text{rs}} = &\frac{\Delta_v^{\text{rs}} (k^{\text{rs}})^2}{\pi C^{\text{rs}}} (v^{\text{rs}}-v_r^{\text{rs}})\\ 
    &+ r^{\text{rs}}(k^{\text{rs}}(2 v^{\text{rs}} - v_r^{\text{rs}} - \bar{v_{\theta}^{\text{rs}}}) - J_{\text{rr}} g^{\text{rs}} s^{\text{rs}}\nonumber\\
    &- J_{\text{rf}} g^{\text{fs}} s^{\text{fs}} - J_{\text{rl}} g^{\text{lts}} s^{\text{lts}}), \nonumber\\
    C^{\text{rs}} \frac{d}{dt} v^{\text{rs}} = &k^{\text{rs}} v^{\text{rs}}(v^{\text{rs}} -v_r^{\text{rs}}-\bar{v_{\theta}^{\text{rs}}})\\ 
    &- \pi C^{\text{rs}} r^{\text{rs}}(\Delta_v^{\text{rs}} + \frac{\pi C^{\text{rs}}}{k^{\text{rs}}} r^{\text{rs}}) + v_r^{\text{rs}} \bar{v_{\theta}^{\text{rs}}} - u^{\text{rs}} + I^{\text{rs}} \nonumber\\
    &+ J_{\text{rr}} g^{\text{rs}} s^{\text{rs}} (E^{\text{rs}}-v^{\text{rs}}) + J_{\text{rf}} g^{\text{fs}} s^{\text{fs}} (E^{\text{fs}}-v^{\text{rs}})\nonumber\\
    &+ J_{\text{rl}} g^{\text{lts}} s^{\text{lts}} (E^{\text{lts}}-v^{\text{rs}}), \nonumber\\
    \tau_u^{\text{rs}} \frac{d}{dt}u^{\text{rs}} = &b^{\text{rs}}(v^{\text{rs}}-v_r^{\text{rs}}) - u^{\text{rs}} + \tau_u^{\text{rs}} \kappa^{\text{rs}} r^{\text{rs}},\\
    \tau_s^{\text{rs}} \frac{d}{dt} s^{\text{rs}} = &-s^{\text{rs}} + \tau_s^{\text{rs}} r^{\text{rs}}, \\
    C^{\text{fs}} \frac{d}{dt}r^{\text{fs}} = &\frac{\Delta_v^{\text{fs}} (k^{\text{fs}})^2}{\pi C^{\text{fs}}} (v^{\text{fs}}-v_r^{\text{fs}})\\ 
    &+ r^{\text{fs}}(k^{\text{fs}}(2 v^{\text{fs}} - v_r^{\text{fs}} - \bar{v_{\theta}^{\text{fs}}}) - J_{\text{fr}} g^{\text{rs}} s^{\text{rs}}\nonumber\\
    &- J_{\text{ff}} g^{\text{fs}} s^{\text{fs}} - J_{\text{fl}} g^{\text{lts}} s^{\text{lts}}), \nonumber\\
    C^{\text{fs}} \frac{d}{dt} v^{\text{fs}} = &k^{\text{fs}} v^{\text{fs}}(v^{\text{fs}} -v_r^{\text{fs}}-\bar{v_{\theta}^{\text{fs}}})\\ 
    &- \pi C^{\text{fs}} r^{\text{fs}}(\Delta_v^{\text{fs}} + \frac{\pi C^{\text{fs}}}{k^{\text{fs}}} r^{\text{fs}}) + v_r^{\text{fs}} \bar{v_{\theta}^{\text{fs}}} - u^{\text{fs}} + I^{\text{fs}} \nonumber\\
    &+ J_{\text{fr}} g^{\text{rs}} s^{\text{rs}} (E^{\text{rs}}-v^{\text{fs}}) + J_{\text{ff}} g^{\text{fs}} s^{\text{fs}} (E^{\text{fs}}-v^{\text{fs}})\nonumber\\
    &+ J_{\text{fl}} g^{\text{lts}} s^{\text{lts}} (E^{\text{lts}}-v^{\text{fs}}), \nonumber\\
    \tau_u^{\text{fs}} \frac{d}{dt}u^{\text{fs}} = &b^{\text{fs}}(v^{\text{fs}}-v_r^{\text{fs}}) - u^{\text{fs}} + \tau_u^{\text{fs}} \kappa^{\text{fs}} r^{\text{fs}},\\
    \tau_s^{\text{fs}} \frac{d}{dt} s^{\text{fs}} = &-s^{\text{fs}} + \tau_s^{\text{fs}} r^{\text{fs}},\\
    C^{\text{lts}} \frac{d}{dt}r^{\text{lts}} = &\frac{\Delta_v^{\text{lts}} (k^{\text{lts}})^2}{\pi C^{\text{lts}}} (v^{\text{lts}}-v_r^{\text{lts}})\\ 
    &+ r^{\text{lts}}(k^{\text{lts}}(2 v^{\text{lts}} - v_r^{\text{lts}} - \bar{v_{\theta}^{\text{lts}}}) - J_{\text{fr}} g^{\text{rs}} s^{\text{rs}}\nonumber\\
    &- J_{\text{lf}} g^{\text{fs}} s^{\text{fs}}), \nonumber\\
    C^{\text{lts}} \frac{d}{dt} v^{\text{lts}} = &k^{\text{lts}} v^{\text{lts}}(v^{\text{lts}} -v_r^{\text{lts}}-\bar{v_{\theta}^{\text{lts}}})\\ 
    &- \pi C^{\text{lts}} r^{\text{lts}}(\Delta_v^{\text{lts}} + \frac{\pi C^{\text{lts}}}{k^{\text{lts}}} r^{\text{lts}}) + v_r^{\text{lts}} \bar{v_{\theta}^{\text{lts}}}\nonumber\\
    &- u^{\text{lts}} + I^{\text{lts}} + J_{\text{lr}} g^{\text{rs}} s^{\text{rs}} (E^{\text{rs}}-v^{\text{lts}})\nonumber\\
    &+ J_{\text{lf}} g^{\text{fs}} s^{\text{fs}} (E^{\text{fs}}-v^{\text{lts}}),\nonumber\\
    \tau_u^{\text{lts}} \frac{d}{dt}u^{\text{lts}} = &b^{\text{lts}}(v^{\text{lts}}-v_r^{\text{lts}}) - u^{\text{lts}} + \tau_u^{\text{lts}} \kappa^{\text{lts}} r^{\text{lts}},\\
    \tau_s^{\text{lts}} \frac{d}{dt} s^{\text{lts}} = &-s^{\text{lts}} + \tau_s^{\text{lts}} r^{\text{lts}}.
\end{align}
\label{eq:eiic_mf}
\end{subequations}

\subsection{Appendix C: Model parameters\label{app:params}}

Here we report the values for all default parameters used in the models studied in this letter.
The default parameters for the regular-spiking neurons given in Tab.\ref{tab:rs} represent a version of the regular-spiking neuron suggested in \cite{izhikevich_dynamical_2007}, modified here to exhibit reduced spike-frequency adaptation and different values of the peak and reset potentials.

\begin{table}[h]
\caption{Regular-spiking neuron parameters\label{tab:rs}}
\begin{ruledtabular}
\begin{tabular}{cccc}
\textrm{Parameter} & \textrm{Value} & \textrm{Parameter} & \textrm{Value}\\
\colrule
$C$ & $100$ & $k$ & $0.7$\\
$v_r$ & $-60$ & $\bar v_{\theta}$ & $-40$\\
$g$ & $1$ & $E$ & $0$\\
$\tau_u$ & $33.33$ & $\tau_s$ & $6.0$\\
$\kappa$ & $10$ & $b$ & $-2.0$\\
$J$ & $15$ & $N$ & $10000$\\
$v_p$ & $1000$ & $v_0$ & $-1000$\\
\end{tabular}
\end{ruledtabular}
\end{table}

The default parameters of the fast-spiking neurons given in Tab.\ref{tab:fs} represent a version of the fast-spiking neuron suggested in \cite{izhikevich_dynamical_2007}, modified here to exhibit different values of the peak and reset potentials.

\begin{table}[H]
\caption{Fast-spiking neuron parameters\label{tab:fs}}
\begin{ruledtabular}
\begin{tabular}{cccc}
\textrm{Parameter} & \textrm{Value} & \textrm{Parameter} & \textrm{Value}\\
\colrule
$C$ & $20$ & $k$ & $1.0$\\
$v_r$ & $-55$ & $\bar v_{\theta}$ & $-40$\\
$g$ & $1$ & $E$ & $-65$\\
$\tau_u$ & $5.0$ & $\tau_s$ & $8.0$\\
$\kappa$ & $0$ & $b$ & $0.025$\\
$J$ & $15$ & $N$ & $10000$\\
$v_p$ & $1000$ & $v_0$ & $-1000$\\
\end{tabular}
\end{ruledtabular}
\end{table}

The default parameters of the low-threshold spiking neurons given in Tab.\ref{tab:lts} represent a version of the low-threshold spiking neuron suggested in \cite{izhikevich_dynamical_2007}, modified here to exhibit different values of the peak and reset potentials.

\begin{table}[H]
\caption{Low-threshold spiking neuron parameters\label{tab:lts}}
\begin{ruledtabular}
\begin{tabular}{cccc}
\textrm{Parameter} & \textrm{Value} & \textrm{Parameter} & \textrm{Value}\\
\colrule
$C$ & $100$ & $k$ & $1.0$\\
$v_r$ & $-56$ & $\bar v_{\theta}$ & $-42$\\
$g$ & $1$ & $E$ & $-65$\\
$\tau_u$ & $33.33$ & $\tau_s$ & $8.0$\\
$\kappa$ & $20$ & $b$ & $8.0$\\
\end{tabular}
\end{ruledtabular}
\end{table}

The synaptic connection strengths used for networks of coupled regular- and fast-spiking neurons are reported in Tab.\ref{tab:two_pop_coupling}; the connection strengths used for networks of coupled regular-, fast-, and low-threshold-spiking neurons are reported in Tab.\ref{tab:three_pop_coupling}

\begin{table}[H]
\caption{Coupling strengths in two-population model\label{tab:two_pop_coupling}}
\begin{ruledtabular}
\begin{tabular}{cccc}
\textrm{Parameter} & \textrm{Value} & \textrm{Parameter} & \textrm{Value}\\
\colrule
$J_{rr}$ & $16$ & $J_{rf}$ & $16$\\
$J_{ff}$ & $4$ & $J_{fr}$ & $4$\\
\end{tabular}
\end{ruledtabular}
\end{table}

\begin{table}[H]
\caption{Coupling strengths in three-population model\label{tab:three_pop_coupling}}
\begin{ruledtabular}
\begin{tabular}{cccc}
\textrm{Parameter} & \textrm{Value} & \textrm{Parameter} & \textrm{Value}\\
\colrule
$J_{rr}$ & $10$ & $J_{rf}$ & $8$\\
$J_{rl}$ & $8$ & $J_{ff}$ & $4$\\
$J_{fr}$ & $8$ & $J_{fl}$ & $4$\\
$J_{lr}$ & $4$ & $J_{lf}$ & $4$\\
\end{tabular}
\end{ruledtabular}
\end{table}

\end{document}